\documentclass[aps,prl,twocolumn,superscriptaddress,byrevtex,showpacs,preprintnumbers,amsmath,amssymb,floatfixm]{revtex4}

\usepackage{wasysym}
\usepackage{graphicx}
\usepackage{dcolumn}
\usepackage{bm}
\usepackage{color}
\begin{document}

\title{Trapping of Rb atoms by ac electric fields}

\author{Sophie Schlunk}
\email{schlunk@fhi-berlin.mpg.de}
\affiliation{Fritz-Haber-Institut der Max-Planck-Gesellschaft,
Faradayweg 4-6,
  14195 Berlin, Germany}
\affiliation{FOM-Institute for Plasmaphysics Rijnhuizen, P.O. Box
1207, 3430 BE Nieuwegein, The Netherlands}
\author{Adela Marian}
\author{Peter Geng}
\affiliation{Fritz-Haber-Institut der Max-Planck-Gesellschaft,
 Faradayweg 4-6, 14195 Berlin, Germany}
\author{Allard P.\ Mosk}
 \altaffiliation[Present address: ]
  {Mesa$^+$ Institute for Nanotechnology, University of Twente, Enschede, The Netherlands}
 \affiliation{FOM-Institute for Plasmaphysics Rijnhuizen, P.O. Box
1207, 3430 BE Nieuwegein, The Netherlands}
\author{Gerard Meijer}
\author{Wieland Sch\"ollkopf}
 \email{wschoell@fhi-berlin.mpg.de}
\affiliation{Fritz-Haber-Institut der Max-Planck-Gesellschaft,
 Faradayweg 4-6, 14195 Berlin, Germany}
\date{\today}

\begin{abstract} We demonstrate trapping of an ultracold gas of
neutral atoms in a macroscopic ac electric trap. Three-dimensional
confinement is obtained by switching between two saddle-point
configurations of the electric field. Stable trapping is observed
in a narrow range of switching frequencies around 60 Hz. The
dynamic confinement of the atoms is directly visualized at
different phases of the ac switching cycle. We observe about
$10^5$ Rb atoms in the 1 mm$^3$ large and several microkelvins deep trap
with a lifetime of approximately 5 s.
\end{abstract}


\maketitle

Trapping neutral particles has proven tremendously successful for
the study of their properties and interactions. Particles in a
low-field seeking quantum state can be trapped in a minimum of a
static field. Particles in a high-field seeking quantum state
cannot be confined in a static field, as static fields cannot
possess a maximum in free space \cite{Wing84,Ketterle92}. Trapping
of atoms and molecules in their lowest-energy quantum level, which
is always high-field seeking, is therefore not possible in static
electric or magnetic fields. These ground-state atoms and
molecules can be confined, however, when electrodynamic fields are
used. Optical dipole traps, for instance, have found widespread
application in ultracold atom experiments \cite{Grimm00}.
Paramagnetic ground-state atoms and molecules can be trapped in ac
magnetic fields, as has been demonstrated for Cs atoms
\cite{Cornell91}.

A more versatile way of trapping ground-state atoms and molecules 
in a large volume is to make use of ac electric fields. To date, different ac electric 
trap configurations have been proposed \cite{Shimizu92,Morinaga94,Peik99}.
The operation principle of an ac electric trap for neutral
particles is analogous to that of the Paul trap for ions
\cite{Paul90}; a potential energy surface is created with a saddle
point at the trap center, resulting in attractive forces
(focusing) in one direction and repulsive forces (defocusing)
along the other two directions. The electric field configuration
is then switched to a second one in which the roles of the forces
are reversed. Alternating between these two configurations at the
appropriate frequency, using either a sine wave or a square wave,
leads to dynamic confinement of the particles.

Trapping with ac electric fields is particularly appealing for polar molecules
that show a first-order Stark effect and that can therefore be
strongly confined. For ground-state ammonia molecules, ac electric 
traps with a depth of several millikelvins have been demonstrated 
\cite{Jacqueline05,Schnell07}, and the motion of the trapped
molecules has been analyzed in detail \cite{Jacqueline06b}. In
addition, any ground-state atom or nonpolar molecule can be
confined in an ac electric trap via their induced dipole moment.
This includes species that are hard to trap otherwise, e.g., singlet homonuclear diatomics. The second-order Stark interaction leads to shallower traps, but
has the advantage that all ground-state sublevels can be trapped.
Recently, about 100 ground-state Sr atoms have been trapped in a
microstructured ac electric trap on a chip with a lifetime of 80
ms \cite{Katori06}.

Here we demonstrate trapping of ground-state $^{87}$Rb atoms in a
macroscopic ac electric trap loaded from a magnetic trap.
Absorption imaging of the cloud of trapped atoms is used to
visualize the dynamics of the confinement. The operation
characteristics of the trap are studied as a function of the
switching frequency. We also discuss the prospects of sympathetic
cooling of molecules with ultracold atoms using spatially
overlapped traps.

\begin{figure}[pt]
\includegraphics[scale=0.62]{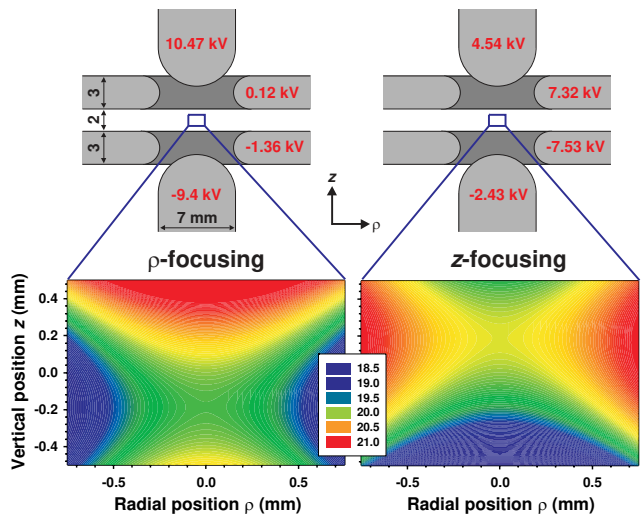}
\caption{Cross section of the ac trap. The two end cap and two ring
electrodes of semicircular shape are arranged for axial symmetry
along the $z$ axis (vertical) and mirror symmetry with respect to
the $z$=0 plane. The distance between the end caps is 6 mm, the
inner diameter of the rings is 6.7 mm, the other dimensions are
indicated in millimeters. The voltage applied is indicated on each
electrode for the $\rho$-focusing (left) and $z$-focusing (right)
configurations. The calculated electric field strengths close to
the trap center are shown in the contour plots (color scale in
kV/cm). The saddle point is vertically displaced due to the
gravity-compensating quadrupolar field component.}
\label{fig:actrap}
\end{figure}

In the experiment, the Rb atoms are collected in a 6-beam
magneto-optical trap (MOT) loaded from a Zeeman slower. After a
short compression of the MOT, optical molasses cooling, and
optical pumping, the atoms are transferred into a spatially
overlapped quadrupole magnetic trap. About $5 \times 10^8$ atoms
in the $F = 2$, $m_F = 2$ hyperfine level at a temperature of 100
$\mu$K are trapped in the magnetic trap which is characterized by
a field gradient of 72 G/cm along its symmetry axis. The
quadrupole magnet is then moved horizontally over 42 cm in 3 s using
a precision translation stage with a nominal position
reproducibility of 3 $\mu$m. The cold cloud is transferred
without significant atom loss or heating \cite{Lewandowski03} into
a second quartz cell. This second ultrahigh vacuum chamber houses
the ac trap. The final position of the cloud is carefully overlapped
with the center of the ac trap. Approximately 1.5 s after arriving
at the final position the magnetic field is switched off. The
magnetic field has completely disappeared after 150 $\mu$s, and
350 $\mu$s later high voltage is applied to the ac trap
electrodes. After a variable trapping time, the trap electrodes
are switched to ground, and the remaining atoms are detected
by absorption imaging.

The ac trap is cylindrically symmetric with respect to the
$z$ axis and consists of four highly polished electrodes of
nonmagnetic stainless steel, as shown in Fig.~\ref{fig:actrap}.
The two end cap electrodes have a hemispherical shape of radius 3.5
mm and are separated by 6 mm. The ring electrodes have an opening
diameter of 6.7 mm, a thickness of 3 mm, and a corresponding inner
semicircular shape of 1.5 mm radius. They are separated by a 2 mm
gap. Care is taken to align the electrodes to the vertical
rotational symmetry axis as well as to the horizontal mirror
symmetry plane.

\begin{figure}[pt]
\includegraphics[scale=0.78]{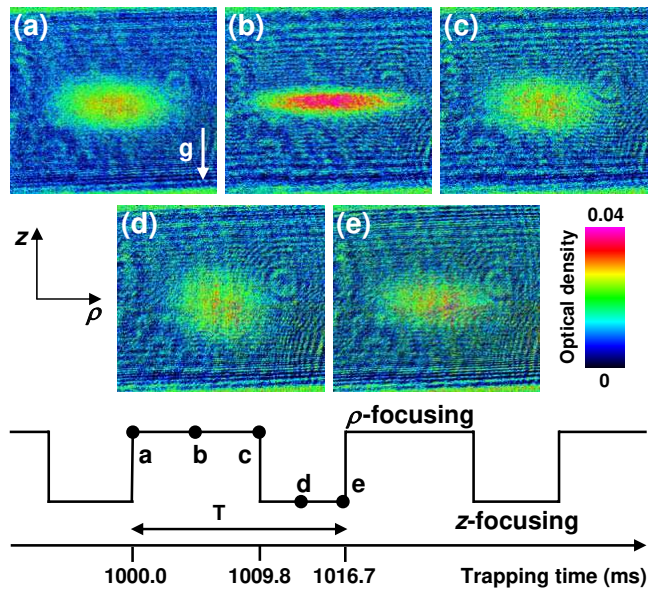}
\caption{Absorption images of the Rb cloud at different times
within the 61st switching cycle, i.e., at the beginning, middle,
and end of the $\rho$-focusing [(a)-(c)] and at the middle and end of
the $z$-focusing [(d),(e)] phase. The corresponding times are
indicated in the switching pattern; the switching
frequency is 60 Hz, giving a period of $T=16.7$ ms. In (b) and (d) the atoms are at the turning points of the micromotion. The image area is $2.2 \times 2$ mm$^2$, such that the field of view covers
exactly the space between the two ring electrodes. Each image is
obtained by averaging 100 pictures and the corresponding optical
density is indicated by the color scale.}
\label{fig:DutyCycleImages}
\end{figure}

The electric field in the trap is a superposition of a dc dipole
field and an ac hexapole field, with an additional dc quadrupole
field to counteract gravity. Our trapping scheme thereby follows
Peik's proposal \cite{Peik99}, and has the advantage that both the
direction and the magnitude of the electric field at the center of
the trap remain constant. Instead of the proposed sine wave,
however, the hexapolar component of the field is switched between
two configurations. The voltages that are applied to the four
electrodes to obtain either one of the electric field
configurations, denoted as $\rho$ focusing and $z$ focusing, are
shown in Fig.~\ref{fig:actrap}. The ac switching frequency is
given by $1/T$, where $T$ is the sum of the durations of the
$\rho$-focusing and $z$-focusing phases. At a given switching
frequency, the relative duration of the two field configurations
can be adjusted. The square-wave alternation between these
configurations as applied in the experiments is schematically
indicated in the lower part of Fig.~\ref{fig:DutyCycleImages}.

The calculated electric field strengths in the trapping region are
plotted in Fig.~\ref{fig:actrap}. At the trap center, the electric
field is 20 kV/cm in both configurations. For the set of voltages
indicated on the left-hand side of Fig.~\ref{fig:actrap}, the
field exhibits a minimum along the $z$ axis and a maximum along
the radial ($\rho$) direction, resulting in forces that are
attractive along $\rho$ and repulsive along $z$. Therefore, we
refer to this state as $\rho$ focusing. For the set of voltages
indicated on the right-hand side of Fig.~\ref{fig:actrap}, the
situation is reversed; the atoms are attracted to the center along
$z$ and repelled in the $\rho$ direction, therefore referred to as
$z$ focusing.

The dynamics of the atoms in the ac trap driven at the switching
frequency, referred to as micromotion in ion traps, is illustrated
in Fig.~\ref{fig:DutyCycleImages}. Five absorption images of the
cloud are shown at various times, labelled (a)$-$(e), within the 61st
trapping cycle. All pictures are taken after 0.1 ms of
time of flight (TOF), reflecting the spatial distribution of the atoms at the switch-off time. All are side-view images of the cloud; i.e., gravity is pointing along the negative $z$ direction. The data are
taken at the optimum frequency of 60 Hz, where the 61st switching
cycle begins after 1000 ms. The switching cycle is asymmetric,
starting with a 59\% $\rho$-focusing phase followed by a 41\%
$z$-focusing phase. Figure \ref{fig:DutyCycleImages}(a) shows the
cloud at the beginning of the cycle where the atoms have just been
$z$ focused. The cloud is slightly oblate.
Figure \ref{fig:DutyCycleImages}(b) is an image of the atoms in the
middle of the $\rho$-focusing phase. From the $z$-focusing phase
the atoms have velocity components along $z$ towards the center of
the trap, and are spreading out along $\rho$. This expansion is slowed 
down, however, by the $\rho$-focusing forces. In the middle of the
$\rho$-focusing phase the atoms are at a turning point of the
micromotion and have a maximum spread in the $\rho$ direction.
Then they are turned around and move back towards the center. In
the $z$ direction, the velocity components of the atoms towards
the center have led to a contraction of the cloud resulting in the
pancake shape shown in Fig.~\ref{fig:DutyCycleImages}(b). As they
are defocused along $z$ during the $\rho$-focusing phase, they
start to spread out in this direction. This can be seen in
Fig.~\ref{fig:DutyCycleImages}(c) which shows the spatial distribution of the atoms at the end
of the $\rho$-focusing phase. The cloud appears to have the same
shape as in Fig.~\ref{fig:DutyCycleImages}(a), but the velocity
components of the atoms are now pointing inwards along $\rho$ and
outwards along $z$. This leads to an approximately spherical shape
in Fig.~\ref{fig:DutyCycleImages}(d) where the cloud has maximum
spread in the $z$ direction. Figure \ref{fig:DutyCycleImages}(e)
shows the cloud at the end of the cycle which is identical in
shape to Fig.~\ref{fig:DutyCycleImages}(a).

\begin{figure}[pt]
\includegraphics[scale=0.5]{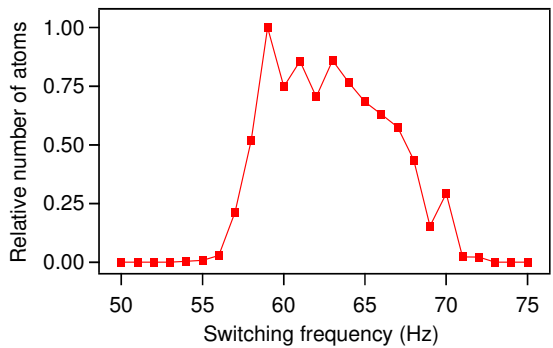}
\caption{Observed number of atoms after 1.0 s of ac trapping as a
function of the switching frequency. Each point is the average of
60 experimental runs. Note that the range of switching frequencies
suitable for trapping is rather narrow.} \label{fig:frequencyscan}
\end{figure}

Figures \ref{fig:DutyCycleImages}(b) and \ref{fig:DutyCycleImages}(d) illustrate why
switching between two saddle-point configurations leads to a net
focusing force. The atoms are always closer to the center when the
defocusing forces are on, and further away while the focusing
forces are on. In Fig.~\ref{fig:DutyCycleImages}(b)
[Fig.~\ref{fig:DutyCycleImages}(d)] the axial [radial] spread of
the atoms is small and therefore the defocusing force is small. On
the other hand, the spread in the focusing $\rho$ [$z$] direction
is big and the focusing forces are accordingly large. Thus, on
average, the micromotion of the atoms leads to bigger focusing
than defocusing forces and therefore the atoms remain confined.
Close to the center of the trap, the force constant along the
$z$ direction is twice as large as the one in the
$\rho$ direction. This difference in forces can be compensated for
by adjusting the asymmetry of the switching cycle. In this
experiment, 59\% of $\rho$ focusing was found to yield the largest
number of trapped atoms, whereas for a symmetric switching cycle
no trapping was observed. A similar behavior was found previously
in the ac trapping of polar molecules
\cite{Jacqueline05,Jacqueline06b}.

\begin{figure}
\includegraphics[scale=0.78]{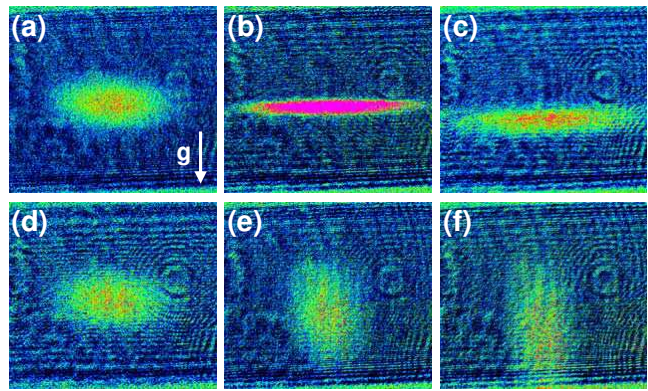}
\caption{Absorption images after different free ballistic
expansion times of 0.1 ms [(a),(d)], 4 ms [(b),(e)], and 7 ms [(c),(f)]. The
ac trap is switched off at the end of the $z$-focusing phase for
(a)-(c) and at the end of the $\rho$-focusing phase for (d)-(f).
For each measurement the trapping time is 1.0 s and the switching
frequency is 60 Hz. All images have the same technical
characteristics as in Fig.~\ref{fig:DutyCycleImages}.}
\label{fig:tofimages}
\end{figure}

The relative number of atoms in the ac trap after 1.0 s of
trapping is plotted in Fig.~\ref{fig:frequencyscan} as a function
of the switching frequency. These data are also taken with 59\% of
$\rho$ focusing. Optimum trapping is observed at a switching
frequency of 60 Hz. At this frequency, we infer from the
absorption images that the number of trapped atoms is about $1
\times 10^5$. After 10 s of ac trapping, the number of atoms
remaining in the trap is reduced to 10\% of this value. This
corresponds to a trap lifetime of about 5 s, presumably limited by
collisions with background gas. As can be seen in
Fig.~\ref{fig:frequencyscan}, the low-frequency cutoff around 56
Hz is followed by a steep increase in the number of atoms to the
maximum at 60 Hz. At higher switching frequencies the number of
atoms decreases smoothly as trapping becomes less efficient. Above
72 Hz no atoms are detected. The range of working frequencies is
narrow compared to that of both Paul traps for ions
\cite{Paul90} and ac traps for polar molecules
\cite{Jacqueline06b}. The observed frequency range is in good
agreement with results from trajectory calculations.

Figure \ref{fig:tofimages} shows the cloud of atoms for a number of
TOF values after a trapping time of 1.0 s  at a switching
frequency of 60 Hz. The upper row of pictures shows the free
ballistic expansion of the atoms directly after the $z$-focusing
phase, whereas for the lower row the ac trap has been switched off
at the end of $\rho$ focusing. In Fig.~\ref{fig:tofimages}(a)
the atoms are imaged after a TOF of 0.1 ms. Given that the atoms have not
moved much since the trap was switched off, this is the atomic
distribution at the end of $z$ focusing also seen in
Fig.~\ref{fig:DutyCycleImages}(a). After 4 ms TOF
[Fig.~\ref{fig:tofimages}(b)], the cloud has a pancake shape as
the velocity distribution along the axial direction leads to
focusing in this direction. Additionally, the cloud is expanding
along $\rho$. Figure \ref{fig:tofimages}(c) is taken after 7 ms of
expansion. The cloud has spread out in the $z$ direction as the
atoms continued to move along $z$; atoms originally at the
top of the cloud, are now at the bottom, and vice versa. In
addition, the atoms are falling towards the electrodes at the
bottom edge of the picture. Still, the cloud has a smaller
diameter in the $z$ direction than in Fig.~\ref{fig:tofimages}(a).
In Figs.~\ref{fig:tofimages}(d)$-$\ref{fig:tofimages}(f) the atoms have just experienced
$\rho$-focusing forces resulting in a different velocity
distribution than in ~\ref{fig:tofimages}(a)$-$\ref{fig:tofimages}(c). Figure \ref{fig:tofimages}(d) shows
the cloud after a TOF of 0.1 ms as already seen in
Fig.~\ref{fig:DutyCycleImages}(c). Now the atoms are accelerating
towards the center in the radial direction and expanding in the
axial direction. In Fig.~\ref{fig:tofimages}(e) after 4 ms  of
free flight the cloud has thus become cigar shaped. In
Fig.~\ref{fig:tofimages}(f) after 7 ms  of TOF, the cloud is even
smaller in the radial direction and is falling out of view.

Free ballistic expansion measurements have also been carried out
to determine the kinetic energy (excluding micromotion) of the
trapped atom cloud. As the observed densities in the ac trap
are on the order of $10^9$ cm$^{-3}$  or less, the cloud is in the
collisionless regime. Therefore, different nonthermal velocity
distributions of the atoms are expected along $\rho$ and $z$. By TOF imaging at the turning points of the micromotion [Figs.~\ref{fig:DutyCycleImages}(b) and ~\ref{fig:DutyCycleImages}(d)], we determine the mean kinetic energy to be
$k_B \times 5 \, \mu$K in the $z$ direction and $k_B \times 20 \,
\mu$K in the $\rho$ direction, with $k_B$ the Boltzmann constant.
Multipole-expansion fits to the calculated electric fields in our trap indicate that higher-order terms up to the decapole are
significant. The finite octupole and decapole components result in
additional forces that considerably reduce the trap depth
\cite{Jacqueline06b}.

The experimental setup presented in this work has been designed
with the aim to produce ultracold molecules via sympathetic
cooling. Cooling of molecules through elastic collisions with
ultracold atoms has been discussed as a promising approach to
bring molecules into the quantum degenerate regime
\cite{Modugno01}. To avoid losses due to inelastic collisions, the
molecules need to be in their ground state \cite{Lara06}.
Therefore, an appealing method would be to confine ground-state
atoms and molecules simultaneously in the same ac trap. However,
polarizable atoms cannot be trapped at the same ac frequencies as
polar molecules because of the largely different Stark interaction
strengths. Thus, an ac trap for polar molecules has to be spatially
overlapped with a different kind of atomic trap, i.e., a magnetic
trap or a dipole trap. Moreover, these traps should not perturb
each other. Here, we have demonstrated that a magnetic trap for
atoms can be spatially overlapped with an ac electric trap. In a
separate experiment, we have also operated this ac trap with
voltages and switching frequencies that would confine ground-state
ND$_3$ molecules, while the atoms remained magnetically trapped.
We have thereby experimentally verified that the Rb atoms in the
magnetic trap are not perturbed by the ac electric fields needed
for trapping molecules. At the same time, the magnetic fields used
to trap the atoms have a negligible effect on the trapping of
closed-shell polar molecules like ND$_3$. The compatibility of
these traps is an important prerequisite for sympathetic cooling
of molecules using magnetically trapped ultracold atoms.

In conclusion, we have demonstrated trapping of about $10^5$ Rb atoms in a macroscopic ac electric trap operated at a switching frequency of 60 Hz. The physical process at the basis of trapping is illustrated with absorption images taken at different phases of the ac switching cycle. The trap depth is several microkelvins and can be increased by applying higher voltages and by using an optimized electrode geometry that reduces higher-order multipole field components. The experimental setup presented here is well suited to pursue cooling of ground-state molecules into the quantum degenerate regime, either via simultaneous trapping of nonpolar molecules and atoms in an ac electric trap, or via overlapping a magnetic trap for atoms and an ac electric trap for polar molecules.

\begin{acknowledgments} This work is part of the research
program of the ``Stichting voor Fundamenteel Onderzoek der Materie
(FOM)". A.\ M.\ would like to thank the Alexander von Humboldt 
Foundation for their support. We
acknowledge fruitful discussions with H.L.\ Bethlem, P.\ L\"utzow,
and S.A.\ Meek, and technical assistance from W.\ Erlebach, G.\
Heyne, and A.J.A.\ van Roij. We thank F.\ Schreck for the data
acquisition software.
\end{acknowledgments}


\begin{thebibliography}{15}
\expandafter\ifx\csname
natexlab\endcsname\relax\def\natexlab#1{#1}\fi
\expandafter\ifx\csname bibnamefont\endcsname\relax
  \def\bibnamefont#1{#1}\fi
\expandafter\ifx\csname bibfnamefont\endcsname\relax
  \def\bibfnamefont#1{#1}\fi
\expandafter\ifx\csname citenamefont\endcsname\relax
  \def\citenamefont#1{#1}\fi
\expandafter\ifx\csname url\endcsname\relax
  \def\url#1{\texttt{#1}}\fi
\expandafter\ifx\csname
urlprefix\endcsname\relax\def\urlprefix{URL }\fi
\providecommand{\bibinfo}[2]{#2}
\providecommand{\eprint}[2][]{\url{#2}}

\bibitem[{\citenamefont{Wing}(1984)}]{Wing84}
\bibinfo{author}{\bibfnamefont{W.~H.} \bibnamefont{Wing}},
  \bibinfo{journal}{Prog. Quant. Electron.} \textbf{\bibinfo{volume}{8}},
  \bibinfo{pages}{181} (\bibinfo{year}{1984}).

\bibitem[{\citenamefont{Ketterle and Pritchard}(1992)}]{Ketterle92}
\bibinfo{author}{\bibfnamefont{W.}~\bibnamefont{Ketterle}} \bibnamefont{and}
  \bibinfo{author}{\bibfnamefont{D.~E.} \bibnamefont{Pritchard}},
  \bibinfo{journal}{Appl. Phys. B} \textbf{\bibinfo{volume}{54}},
  \bibinfo{pages}{403} (\bibinfo{year}{1992}).

\bibitem[{\citenamefont{Grimm et~al.}(2000)\citenamefont{Grimm, Weidem\"uller,
  and Ovchinnikov}}]{Grimm00}
\bibinfo{author}{\bibfnamefont{R.}~\bibnamefont{Grimm}},
  \bibinfo{author}{\bibfnamefont{M.}~\bibnamefont{Weidem\"uller}},
  \bibnamefont{and} \bibinfo{author}{\bibfnamefont{Y.~B.}
  \bibnamefont{Ovchinnikov}},     
  \bibinfo{journal}{Adv. At., Mol., Opt. Phys.}
  \textbf{\bibinfo{volume}{42}},
  \bibinfo{pages}{95} (\bibinfo{year}{2000}).

\bibitem[{\citenamefont{Cornell et~al.}(1991)\citenamefont{Cornell, Monroe, and
  Wieman}}]{Cornell91}
\bibinfo{author}{\bibfnamefont{E.~A.} \bibnamefont{Cornell}},
  \bibinfo{author}{\bibfnamefont{C.}~\bibnamefont{Monroe}}, \bibnamefont{and}
  \bibinfo{author}{\bibfnamefont{C.~E.} \bibnamefont{Wieman}},
  \bibinfo{journal}{Phys. Rev. Lett.} \textbf{\bibinfo{volume}{67}},
  \bibinfo{pages}{2439} (\bibinfo{year}{1991}).

\bibitem[{\citenamefont{Shimizu and Morinaga}(1992)}]{Shimizu92}
\bibinfo{author}{\bibfnamefont{F.}~\bibnamefont{Shimizu}} \bibnamefont{and}
  \bibinfo{author}{\bibfnamefont{M.}~\bibnamefont{Morinaga}},
  \bibinfo{journal}{Jpn. J. Appl. Phys.} \textbf{\bibinfo{volume}{31}},
  \bibinfo{pages}{L1721} (\bibinfo{year}{1992}).

\bibitem[{\citenamefont{Morinaga and Shimizu}(1994)}]{Morinaga94}
\bibinfo{author}{\bibfnamefont{M.}~\bibnamefont{Morinaga}} \bibnamefont{and}
  \bibinfo{author}{\bibfnamefont{F.}~\bibnamefont{Shimizu}},
  \bibinfo{journal}{Laser Phys.} \textbf{\bibinfo{volume}{4}},
  \bibinfo{pages}{412} (\bibinfo{year}{1994}).

\bibitem[{\citenamefont{Peik}(1999)}]{Peik99}
\bibinfo{author}{\bibfnamefont{E.}~\bibnamefont{Peik}}, \bibinfo{journal}{Eur.
  Phys. J. D} \textbf{\bibinfo{volume}{6}}, \bibinfo{pages}{179}
  (\bibinfo{year}{1999}).

\bibitem[{\citenamefont{Paul}(1990)}]{Paul90}
\bibinfo{author}{\bibfnamefont{W.}~\bibnamefont{Paul}}, \bibinfo{journal}{Rev.
  Mod. Phys.} \textbf{\bibinfo{volume}{62}}, \bibinfo{pages}{531}
  (\bibinfo{year}{1990}).

\bibitem[{\citenamefont{van Veldhoven et~al.}(2005)\citenamefont{van Veldhoven,
  Bethlem, and Meijer}}]{Jacqueline05}
\bibinfo{author}{\bibfnamefont{J.}~\bibnamefont{van Veldhoven}},
  \bibinfo{author}{\bibfnamefont{H.~L.} \bibnamefont{Bethlem}},
  \bibnamefont{and} \bibinfo{author}{\bibfnamefont{G.}~\bibnamefont{Meijer}},
  \bibinfo{journal}{Phys. Rev. Lett.} \textbf{\bibinfo{volume}{94}},
  \bibinfo{pages}{083001} (\bibinfo{year}{2005}).

\bibitem[{\citenamefont{Schnell et~al.}(2007)\citenamefont{Schnell}}]{Schnell07}
\bibinfo{author}{\bibfnamefont{M.}~\bibnamefont{Schnell}}
\emph{et al}.,
   \bibinfo{journal}{J. Phys. Chem. A} (to be published).

\bibitem[{\citenamefont{Bethlem et~al.}(2006)\citenamefont{Bethlem, van
  Veldhoven, Schnell, and Meijer}}]{Jacqueline06b}
\bibinfo{author}{\bibfnamefont{H.~L.} \bibnamefont{Bethlem}}
 \emph{et al}.,
   \bibinfo{journal}{Phys. Rev. A} \textbf{\bibinfo{volume}{74}},
  \bibinfo{pages}{063403} (\bibinfo{year}{2006}).

\bibitem[{\citenamefont{Kishimoto et~al.}(2006)\citenamefont{Kishimoto,
  Hachisu, Fujiki, Nagato, Yasuda, and Katori}}]{Katori06}
\bibinfo{author}{\bibfnamefont{T.}~\bibnamefont{Kishimoto}}
\emph{et al}.,
  \bibinfo{journal}{Phys. Rev. Lett.} \textbf{\bibinfo{volume}{96}},
  \bibinfo{pages}{123001} (\bibinfo{year}{2006}).

\bibitem[{\citenamefont{Lewandowski et~al.}(2003)\citenamefont{Lewandowski,
  Harber, Whitaker, and Cornell}}]{Lewandowski03}
\bibinfo{author}{\bibfnamefont{H.~J.} \bibnamefont{Lewandowski}}
\emph{et al}.,
\bibinfo{journal}{J. Low Temp. Phys.}
  \textbf{\bibinfo{volume}{132}}, \bibinfo{pages}{309} (\bibinfo{year}{2003}).

\bibitem[{\citenamefont{Modugno et~al.}(2001)\citenamefont{Modugno, Ferrari,
  Roati, Brecha, Simoni, and Inguscio}}]{Modugno01}
\bibinfo{author}{\bibfnamefont{G.}~\bibnamefont{Modugno}}
\emph{et al}.,
  \bibinfo{journal}{Science} \textbf{\bibinfo{volume}{294}},
  \bibinfo{pages}{1320} (\bibinfo{year}{2001}).

\bibitem[{\citenamefont{Lara et~al.}(2006)\citenamefont{Lara, Bohn, Potter,
  Soldan, and Hutson}}]{Lara06}
\bibinfo{author}{\bibfnamefont{M.}~\bibnamefont{Lara}}
\emph{et al}.,
  \bibinfo{journal}{Phys. Rev. Lett.} \textbf{\bibinfo{volume}{97}},
  \bibinfo{pages}{183201} (\bibinfo{year}{2006}).

\end{thebibliography}
\end{document}